\documentclass[%
 reprint,
 superscriptaddress,
 amsmath,amssymb,
 aps,
 pra,
showpacs
]{revtex4-1}

\setcitestyle{round,authoryear}
\usepackage{color}
\usepackage{multirow}
\usepackage{graphicx}
\usepackage{dcolumn}
\usepackage{bm}
\usepackage[mathlines]{lineno}
\usepackage{ifthen}
\usepackage{layouts}

\newcommand{\note}[1]{{\small {\bf \color{red} #1}}} 
\newcommand*{\figref}[1]{Figure~\ref{#1}}
\newcommand*{\eqnref}[1]{\ifthenelse{\equal{#1}{}}{\note{EQUATION}}{Eq.~\eqref{#1}}}
\newcommand*{\secref}[1]{\S\ref{#1}}
\newcommand*{\tabref}[1]{Table~\ref{#1}}
\renewcommand{\vec}[1]{\ensuremath{\boldsymbol{#1}}}
\newcommand*{\mat}[1]{\ensuremath{\boldsymbol{\mathsf{#1}}}}
\newcommand*{\set}[1]{\ensuremath{\boldsymbol{\mathcal{#1}}}}
\newcommand*{\mi}[1]{\ensuremath{\boldsymbol{\mathsf{#1}}}}

\newcommand*{\xmin}[1][]{\ensuremath{\vec x^{#1}_\mathrm{min}}}
\newcommand*{\xmax}[1][]{\ensuremath{\vec x^{#1}_\mathrm{max}}}

\newcommand*{\program}[1]{\textsc{#1}}
\newcommand*{\polaris}{\mbox{\program{Polaris(MD)}}}
\newcommand*{\Python}{\program{Python}}
\newcommand*{\SymPy}{\program{SymPy}}

\newcommand*{\T}{\ensuremath{\mathcal{T}}}

\newcommand\plotone[1]{%
 \centering
 \leavevmode
 \includegraphics[width={\columnwidth}]{#1}%
 \typeout{PlotOne: \the\columnwidth}
}%
\newcommand\plottwo[2]{{%
 \columnwidth=.5\columnwidth
 \includegraphics[width={\columnwidth}]{#1}%
 \hfil
 \includegraphics[width={\columnwidth}]{#2}%
 \typeout{PlotTwo: \the\columnwidth}
}}%
\newcommand\plotthree[3]{{%
 \centering
 \leavevmode
 \columnwidth=.30\textwidth
 \includegraphics[width={\columnwidth}]{#1}%
 \hfil
 \includegraphics[width={\columnwidth}]{#2}%
 \hfil
 \includegraphics[width={\columnwidth}]{#3}%
 \typeout{PlotThree: \the\columnwidth}
}}%

\begin{document}
\ifdefined\DIFadd
\newcolumntype{d}{c}
\fi

\preprint{APS/123-QED}

\title{The Fast Multipole Method and Point Dipole Moment Polarizable Force Fields }%


\author{Jonathan P. Coles}%
 \email{jonathan.coles@exascale-computing.eu}
 \affiliation{%
    Exascale Research Computing Lab, Campus Teratec, 2 Rue de la Piquetterie, 91680 Bruyeres-le-Chatel, France
}%
\author{Michel Masella}%
 \email{michel.masella@cea.fr}
 \affiliation{%
    Laboratoire de Biologie Structurale et Radiobiologie, Service de Bio\'energétique Biologie Structurale et M\'ecanismes, 
    Institut de biologie et de technologies de Saclay, CEA Saclay, F-91191 Gif sur Yvette Cedex, France
}%

\date{\today}

\begin{abstract}
    We present an implementation of the fast multipole method for computing
    coulombic electrostatic and polarization forces from polarizable
    force-fields based on induced point dipole moments. We demonstrate the
    expected $O(N)$ scaling of that approach by performing single energy point
    calculations on hexamer protein subunits of the mature HIV-1 capsid. We
    also show the long time energy conservation in molecular dynamics at the
    nanosecond scale by performing simulations of a protein complex embedded in
    a coarse-grained solvent using a standard integrator and a multiple
    time step integrator. Our tests show the applicability of FMM combined with
    state-of-the-art chemical models in molecular dynamical systems.
\end{abstract}

\maketitle

\section{\label{sec:Introduction} Introduction}

In $N$-body simulations, the long-range potentials, such as gravitational or
electrostatic potentials, pose the greatest computational difficulty as their
cost scales with the number of particles as $O(N^2)$. Reducing this complexity
has been the subject of intense study for over 40 years and has seen the
development of many successful algorithms. Particularly important have been the
Ewald summation scheme, which achieved a scaling of $O(N^{3/2})$
\citep{perram88}, and its extension,  the particle-mesh-Ewald (PME) scheme
\citep{1993JChPh..9810089D} based on fast Fourier transforms (FFTs), which
scales as $O(N \log N)$. The tree-code from 
\citet{1986Natur.324..446B} also reduced the cost to $O(N \log N)$ using
multipole expansions of the potential.  From a purely algorithmic perspective,
the most performant algorithm should be the fast multipole method (FMM)
proposed by \citet{1987JCoPh..73..325G}, which scales as
$O(N)$.  However, despite volumes of theoretical work, the FMM, on which we
focus here, has not been widely used in production molecular dynamics (MD)
simulations due to the perceived complexity to implement it, hidden constants
that affect its scaling, issues with multiple time step integrators, and early
concerns that energy cannot be conserved well enough unless a prohibitively
high accuracy is used \citep{JCC:JCC7,1997JCP}.  For these reasons, PME is
presently very popular among many molecular dynamics codes
\citep{doi:10.1146/annurev.biophys.28.1.155} to handle periodic molecular
systems containing up to millions of atoms \citep{zhaonature2013}, even if FFT
does not scale well on modern supercomputer architectures due to high
communication overhead. 

Interest in FMM has recently resurfaced again because of these scaling problems
and the desire to solve much larger systems
\citep{PhysRevE.88.063308,Yokota:2012:TSF:2403996.2403998}.  Several successful
implementations have existed in the astrophysics community for many years to
evolve the extremely large number of particles typically representing dark
matter \citep[e.g.,][]{2008Natur.454..735D,2009MNRAS.398L..21S}. In this paper
we present an implementation of FMM to perform molecular dynamics simulations based on
advanced many-body interatomic potentials (force-fields) including polarization
effects described according to the induced point dipole approach. 

The development of polarizable force-fields represents an important step in 
molecular modeling, as such fields have been shown to greatly improve the
description of complex microscopic systems, such as physical interfaces and
those involving charged species
\citep{dang00,lopes09,cieplak09,chang06,jungwirth06,real13,salanne11}.
Polarization is also suspected to greatly improve the description of amino-acid
interactions, which will lead to further promising computational techniques
devoted to theoretically investigations of protein folding \citep{scarpazza13}. 
    
Among the polarizable force-fields proposed so far, the second author has
developed a particular kind of force-field which couples easily and efficiently
with a polarizable coarse grained approach to model the solvent
\citep{masella08,masella11}. Here, a complex solute (e.g., a protein) is
modelled at the atomic level while the solvent is modeled as a set of
polarizable pseudo-particles. The main feature of this approach is the
systematic cut off of the solute-solvent interactions whose computation scales
as $O(N)$ (with $N$ the atomic solute size), while no cut off is considered in
computing intra-solute interactions. In other words, the computation of
intra-solute electrostatic interactions corresponds to an electrostatic free
boundary condition problem. Hence, coupling the modeling approach and an FMM
scheme to compute the intra-solute interactions will lead to a full $O(N)$
polarizable approach to investigate the solvation of large and complex solutes,
without applying any cut-off to intra-solute interactions. Note that even if
the solute-solvent interactions are truncated for distances larger than a
reference one (typically 12 \AA), we recently proposed a multi-scale, coarse
grained approach, which allows the solute-solvent truncation drawbacks to be
minimized, while still maintaining the $O(N)$ complexity of the solute-solvent
interaction computational scheme \citep{masella13}.

Our FMM method is based on the work of \citet{2002JCoPh.179...27D} that
uses a binary tree spatial decomposition and conserves linear momentum by
construction (via a dual tree walk approach to compute multipole interactions).
Accuracy is controlled by an opening angle tolerance parameter $\theta$ at
fixed multipole expansion order.  

In this paper we first review in \secref{sec:Theory} the induced point dipole
method and the mathematical basis of FMM.  In \secref{sec:Implementation} we
discuss some technical details of our implementation and in \secref{sec:tests}
we present tests of the accuracy and precision, particularly when coupling FMM
with a multiple time steps integrator devoted to induced dipole moment based
force-fields. We also discuss the efficiency of an FMM approach compared
to PME when simulating a large molecular system and accounting for
polarization. We present our conclusions in \secref{sec:Conclusions}.

\section{\label{sec:Theory} Theory}
\subsection{\label{sec:Force fields} Induced point dipole polarizable force-fields}

Molecular modeling approaches consider the total interatomic potential $U$ as a
sum of different energy terms. For polarizable force-fields, a sum of four
terms is commonly considered
\begin{equation}
    U = U^\mathrm{short} + U^\mathrm{rel} + U^{qq'} + U^\mathrm{pol}
\end{equation}
Here, $U^\mathrm{short}$ is a short range energy term describing the short
range atom-atom repulsive effects and, usually, dispersion effects. The term is
normally additive in nature, however, there have been attempts to consider
short range many-body potentials to account for short range electronic cloud
reorganization effects \citep[e.g.,][]{real13}. $U^\mathrm{rel}$ is
the sum of the common stretching, bending and torsional potentials, which
describe the interactions among covalently bonded atoms.

The standard pairwise Coulomb potential $U^{qq'}$ is based on static charges
centered on the system atoms
\begin{equation} \label{equ:Uqq}
    U^{qq'} = \frac12 \sum_{i} \Phi^q_i = \dfrac{1}{8\pi \epsilon_0}\sum_{i} q_i \sum^{*}_{j} {q_j \phi(\vec x_i - \vec x_j)}
\end{equation}
where the Green's function is $\phi(\vec x_i - \vec x_j) = {|\vec x_i - \vec
x_j|^{-1}}$.  The superscript $^*$ denotes sums for which a subset
of atoms is excluded (commonly, the atoms $j$ separated by less than two chemical bonds
from an atom $i$).

$U^\mathrm{pol}$ is the polarization term. When considering an induced dipole
moment polarization approach, a set of additional degrees of freedom is
introduced, the induced dipole moments $\vec{\mu}_i$, that obey 
\begin{equation}
    \vec {\mu}_i = \mat \alpha_i \cdot \left( \vec {E}_i^q + \vec E^\mu_i \right)
    \label{eq:dipole moment}
\end{equation}
Here, $\mat \alpha_i$ is the polarizability tensor of the polarizable atom $i$,
$\vec{E}_i^q$ and $\vec E^\mu_i$ are the electric fields generated on atom $i$
by the surrounding static charges $q_j$ and the surrounding induced dipoles
$\vec {\mu}_j$, respectively. These two electric fields are defined as
\begin{equation}
    \vec E^q_i = -\sum_{j}^{*} q_j\ \vec\nabla\phi(\vec x_i - \vec x_j)
\end{equation}
and
\begin{equation}
    \vec E^\mu_i = -\sum_j^{*} \vec\mu_j \cdot \mat T_{ij}
\end{equation}
The dipolar tensor $\mat T_{ij}$ is the second derivative of the Green's function
\begin{equation}
    \mat T_{ij} = \vec\nabla^2\phi(\vec r = \vec x_i - \vec x_j)%
                = \frac{1}{r^3}\left({\mat 1} - {{3 \vec r \otimes \vec r} \over {r^2}} \right)
\end{equation}
where $\otimes$ is the tensor outer product. $\mat T_{ij}$ is, however, usually
altered to account for interatomic short range damping effects. Following
the original ideas of \citet{1981CP.....59..341T}, this is achieved
by adding to $\mat T_{ij}$ a specific short range dipolar tensor $\mat
T_{ij}^\mathrm{damp}$.
 
The interactions of the dipoles with the above two electric fields give rise to
two additional potentials, namely
\begin{equation}
    \Phi^{\mu q}_i = \vec\mu_i \cdot \vec E^q_i
        \quad\mathrm{and}\quad 
    \Phi^{\mu\mu}_i = \vec\mu_i \cdot \vec E^\mu_i
\end{equation}
 
$U^\mathrm{pol}$ also typically accounts for a third energy term, $A(\vec\mu)$,
quantifying the energy cost to create a dipole $\vec{\mu}$. In standard induced
dipole moment implementations, this third term leads to the fundamental
relationship
\begin{equation} \label{eq:upol_derivatives}
    {\partial U^\mathrm{pol} \over \partial \vec\mu} = 0
\end{equation}
For instance, when the dipoles obey the linear relation \eqref{eq:dipole moment}
and the polarizability tensors $\mat{\alpha}_i$ are taken as isotopic
polarizabilities (i.e., they are a scalar quantity), the above
condition is met by taking  
\begin{equation}
    A(\vec \mu) = \frac12 \sum_i {\ |\vec\mu_i|^2 \over \vec\alpha_i}
\end{equation}
Note that alternative induced dipole moment approaches where the dipoles obey
more complex relations have been proposed \citep[e.g.,][]{haduong02}.
In that case, a specific $A(\vec \mu) $ energy term is considered that fulfills
the relation~\eqref{eq:upol_derivatives}.  Regardless of the form of
$A(\vec\mu)$, the total dipole potential energy nevertheless obeys
\begin{equation}
    U^\mathrm{pol}= A(\vec\mu) - \sum_{i} \Phi^{\mu q}_i - \frac12 \sum_i \Phi^{\mu\mu}_i
\end{equation}
%

\subsection{\label{sec:FMM} The Fast Multipole Method}

The fast multipole method (FMM) \citep{1987JCoPh..73..325G} is a technique for
approximating a long range potential
\begin{equation}
    \Phi(\vec x_b) = \sum_{a\in \set P} q_a \phi(\vec x_b - \vec x_a)
\end{equation}
of a set of particles $\set P$ via a multipole
expansion of the Green's function $\phi$.  Typically, the particles are
organized by a hierarchical spatial decomposition such as an oct-tree or binary
tree. We consider the interaction of the multipole expansions of the
particles in pairs of tree nodes $A,B$ centered at $\vec z_A, \vec z_B$,
and containing particles labeled $a$ and $b$, respectively. 

The Taylor expansion of $\phi$ to $p^\mathrm{th}$ order about
both centers in Cartesian coordinates yields the following expression
\begin{equation}
    \phi(\vec x_b - \vec x_a) \approx
        \sum_{|\mi n| \leq p} 
        \sum_{|\mi m| \leq p-|\mi n|}
        \frac{(-1)^{|\mi n|}}{\mi n! \mi m!}\vec r_b^{\mi n}\vec r_a^{\mi m} \vec\nabla^{\mi n + \mi m}\phi(\vec z_B - \vec z_A)
\end{equation}
where $\vec r_b = \vec x_b - \vec z_B$ and $\vec r_a = \vec x_a - \vec z_A$. Here
we make use of multi-index notation, whose relevant properties are summarized in \secref{sec:notation}.
 
Grouping terms, we can express the multipole expansion for a node $A$ (or $B$) as 
\begin{equation}
\label{eq:multipole}
    M_{\mi n}(\vec z_A) = \sum_{a \in A} q_a \frac{(-1)^{|\mi n|}}{\mi n!}\vec r_a^{\mi n}
\end{equation}
Multipoles can be computed efficiently for all parent nodes by combining
the multipoles of child nodes using the shifting operator
\begin{equation}
    M_{\mi n}(\vec z + \vec x) = \sum_{|\mi k| \leq |\mi n|} \frac{\vec x^{\mi k}}{\mi k!} M_{\mi n - \mi k}(\vec z)
\end{equation}
If two nodes are sufficiently distant for the expansions to be accurate---as
determined by an implementation specific multipole acceptance criteria (MAC)
which we will discuss later---the multipoles may be transformed into a local
expansion (or field tensor) about another point, e.g.,
\begin{equation}
    F_{\mi n}(\vec z_B) = \sum_{|\mi m| \leq p-|\mi n|} M_{\mi m}(\vec z_A) \vec\nabla^{\mi n + \mi m}\phi(\vec z_B - \vec z_A)
\end{equation}
Performing this evaluation on both $A$ and $B$ symmetrically naturally
satisfies Newton's third law.

The final evaluation of the potential proceeds as follows. The field tensors
from node-node interactions are accumulated from parent to child down the tree
by the shifting formula
\begin{equation}
    F_{\mi n}(\vec z + \vec x) = \sum_{|\mi k| \leq p-|\mi n|} \frac{\vec x^{\mi k}}{\mi k!} F_{\mi n + \mi k}(\vec z)
\end{equation}
such that $F_{\mi n}$ in the leaf nodes will be the sum of all the field tensors of all
parent nodes and any node-node interactions it had itself.
The approximated potential (or any $k^\mathrm{th}$ order derivative) at the positions of
particles in a leaf node $B$ is then
\begin{equation}
    \vec\nabla^k\Phi(\vec x_b) \approx q_b \sum_{|\mi n| \leq p-k} \frac{1}{\mi n!}(\vec x_b - \vec z_B)^{\mi n} F_{\mi n + k}(\vec z_B)
\end{equation}
%

\subsection{\label{FMM and point induced dipole moment} FMM and induced point dipole moments}

An induced point dipole $\vec\mu_i$ can be viewed as resulting from a pair
of particles with charges $\pm q^\mu_i$ located at a distance $|\delta \vec
l_i|$ from the dipole center such that
\begin{equation}
    \vec\mu_i = 2 |q^\mu_i| \delta \vec l_i
\end{equation}
From the above relation, it is straightforward to compute the polarization
energy and forces using FMM. With the dipole charge set $\{q^\mu_i \}$ and by
considering  $\vec x_i^\pm = \vec x_i \pm \delta \vec l_i$,  we can
reformulate the potentials in $U^\mathrm{pol}$ such that they take a similar
form as the standard Coulomb potential
\begin{align}
    \Phi^{\mu q}_i &= \vec\mu_i \cdot \vec E^q_i \nonumber \\
     &= q^\mu_i \sum_j^* q_j \left( \phi(\vec x_i^+ - \vec x_j ) -  \phi(\vec x_i^-  - \vec x_j) \right)
\label{Phi mu E}
\end{align}
and 
\begin{align}
    \Phi^{\mu\mu}_i &= \vec\mu_i \cdot \vec E^\mu_i \nonumber \\
    &= q^\mu_i \sum_j^* q^\mu_j \left[ \left( \phi(\vec x_i^+ - \vec x_j^+) - \phi(\vec x_i^+ - \vec x_j^-) \right) \right. \nonumber \\
    &\qquad \qquad \;\; \left. - \left( \phi(\vec x_i^- - \vec x_j^+) - \phi(\vec x_i^- - \vec x_j^-) \right) \right]
\label{Phi mu mu}
\end{align}
    
Note that if $U^\mathrm{pol}$ obeys \eqnref{eq:upol_derivatives}, then the
derivatives of $U^\mathrm{pol}$ with respect to the charges $\{q^\mu_i\}$ and the
vectors $\delta \vec l_i$ vanish.  Hence, all the formulas discussed in the
above section can be used as such to compute the polarization energy and
forces. The same $p^\mathrm{th}$ order expansion is used to estimate
equations \eqref{Phi mu E} and \eqref{Phi mu mu} as for the Coulomb potential despite
$\vec\mu_i$ being related to a second order derivative of the Green's function. Our
later tests suggest this has no effect on the resulting precision of the
simulations.

Moreover, the electric fields in equation \eqref{eq:dipole moment}, which the
induced dipole moments obey, are also computable using FMM.  The static
electric field $\vec E_i^q$ can be computed by considering the spatial
derivatives of the electrostatic potential acting on atom $i$ and generated by
the surrounding charges $q_j$. Likewise, the dynamical electric field $\vec
E_i^\mu$ can be computed by considering the spatial derivatives of the
electrostatic potential generated on atom $i$ by the surrounding set of charges
$q^\mu_j$. 

\section{Implementation} \label{sec:Implementation}

Here we will describe briefly our implementation of the FMM in our own
molecular dynamics code \polaris{} \citep{masella08,masella11,JCC:JCC23237}.
The implementation is similar to that described in \citet{2002JCoPh.179...27D},
but we account for the Coulomb, dipole-static electric field, and dipole-dipole
interactions.  Moreover, we also account for the specificities of the
polarizable force-fields implemented in the code \polaris{} by allowing only a
subset of atoms to generate the static electric field acting on a second subset
of polarizable atoms (both these subsets of atoms can be different, and the
charges $q^E_i$ generating the electric field can also be different from the
Coulomb $q_i$ ones \citep{masella03}).  The atomic
polarizability tensor $\mat \alpha_i$ is taken to be isotropic and is replaced
by a scalar $\alpha_i$. 

Regardless of the charge set generating a potential, the approximate
potential~$\Phi$ consists of two components
\begin{equation}
    \Phi = \Phi^\mathrm{direct} + \Phi^\mathrm{fmm}
\end{equation}
The first component is computed via direct particle-particle interactions while
the second is computed via FMM. The direct component includes particles that
are too close for the multipole expansion to be accurate enough and also for
interactions that are cheaper to compute directly than with the expansions.
Other effects, such as damping the polarization effects at short range and the
atom-atom repulsive potential $U^\mathrm{rep}$ are also computed directly using
a list of neighbors when a cell has a ``self''-interaction. We explicitly
consider the induced dipole moments \{$\mat \mu_i$\} for direct
particle-particle interactions and not the $q^\mu_i$ and $\delta \vec l_i$
quantities.

The atoms are organized via an adaptive kd-tree spatial decomposition. A
binary tree, as opposed to an oct-tree, is particularly well suited for
non-uniformly distributed structures, or structures that are not roughly cubic
in extent.  This is an important feature for simulating a complex solute
solvated within a coarse-grained solvent box (see below).  For efficiency, we
build one tree with all the atoms $\T^\mathrm{atom}$ and a second tree with
only the polarizable atoms $\T^\mu$.

The tree is constructed by recursively splitting the bounding volume across the
longest dimension at the location of the geometric center of the particles
within each cell.  The recursion stops once a cell has no more than
$N_\mathrm{bucket}$ particles, where $N_\mathrm{bucket}$ is a parameter we set
to 8. Larger values decrease the size of the tree at the expense of increased
direct sum work. The multipole expansions are computed for all leaf nodes
and then combined up the tree to the root node. For cells with dipole sites,
the two-particle approximation discussed in \secref{FMM and point induced dipole moment}
is computed on the fly and the contribution of the two particles is added to the 
multipole. The two particles are never considered as real particles in the simulation
and are not part of the domain decomposition. This avoids a potential problem whereby
a cell boundary might divide a dipole pair.

In addition to typical tree book keeping data, each cell stores information for
each of the three categories of particles: all atoms, atoms generating $\vec
E^q$, and polarizable atoms. This information includes the geometric center
$\vec z$ of the particle positions; the bounds of the tightest enclosing box
$\xmin, \xmax$; the distance $r_\mathrm{enc}$ from $\vec z$ to the most
distant corner of the bounding box; the multipole coefficients $M_{\mi n}$; and
the field tensor coefficients $F_{\mi n}$. For cells with less than 3
particles, we add a small padding to avoid planar or singular box sizes. The
choice of radius $r_\mathrm{enc}$ allows parent cell sizes and positions to be
easily derived from the children.  Each cell requires approximately
$N_\mathrm{cellsize}=3200$ bytes. Hence, assuming a perfectly balanced tree, a
simulation with $N=10^6$ particles would currently only require
$2N/N_\mathrm{bucket} \times N_\mathrm{cellsize} \sim 810$ MB to store the
tree.

We control the accuracy of the multipole calculation by fixing the expansion
order $p$ of the multipoles and defining the user adjustable parameter $\theta
\in [0,1)$.  Nodes $A,B$ are allowed to interact via their multipoles if they
satisfy the multipole acceptance criteria (MAC)
\begin{equation}
    \label{eq:mac}
    \theta \cdot |\vec z_A - \vec z_B| > (r_{\mathrm{enc},A} + r_{\mathrm{enc},B})
\end{equation}
When $\theta=1$ the enclosing spheres of the nodes are allowed to be exactly
touching, while when $\theta=0$ the nodes are never allowed to interact via
their multipoles and all interactions are computed using a direct sum.  The FMM
of \citet{1987JCoPh..73..325G} controls accuracy in a similar manner with the
expansion order $p$ of the multipoles and a well-separated parameter $ws$ that
determines the minimum distance for a multipole interaction. 

Very recent work by \citet{2014ComAC...1....1D} suggests that at
$\theta=1$ only small improvements to the precision are gained by large
increases in $p$, while at $\theta < 1$ large gains in precision are possible
for small changes in $p$. In \secref{sec:tests} we explore a range of $\theta$
values while fixing the expansion order at $p=5$.

Interactions are determined by descending the tree in a dual-walk fashion
whereby we maintain a stack of interaction pairs.  The walk begins by placing
nodes $A,B$ equal to the root node on the stack.  At each iteration of the
walk, the top most nodes are removed. A direct interaction between the nodes is
considered if the total number of particles is less than $\sim$64, or the nodes
are both leaves. If a direct interaction is not performed but the nodes satisfy
the MAC, a multipole interaction is computed. The field tensors are accumulated
for the mutual interaction of the nodes, thus ensuring conservation of
momentum.  If the MAC is not satisfied, interactions with the children of the
node with the larger $r_\mathrm{enc}$ are placed on the stack. If the larger
node happens to be a leaf, then the children of the smaller node are taken
instead.  By way of example, if node $B$ was larger, the interactions
$AB_\mathrm{left}$, $AB_\mathrm{right}$, and $B_\mathrm{left} B_\mathrm{right}$
are placed on the stack. When the two nodes are equal and a self interaction is
not possible then all unique combinations child interactions are placed on the
stack.  The procedure ends when there is no more work left on the stack.

Calculating all the long range forces involves several applications of the
above procedure applied in three stages. We first calculate the coulomb forces
$\vec\nabla\Phi^q$ and static electric field $\vec E^q$ using
$\T^\mathrm{atom}$. All particles with non-zero $q_i$ interact with each other,
but only those with non-zero $q^E_i$ are allowed to contribute to the electric
field. The electric field is only evaluated at the positions of polarizable
particles (those with non-zero $\vec\mu_i$). Second, we iteratively solve for
the dipoles $\vec\mu_i = \alpha_i \cdot \vec E_i$ using $\T^\mu$.  At each
iteration we recalculate the moments and perform an FMM interaction tree walk.
The details of the iterative procedure have been described earlier in
\citet{masella08}. This has been shown to allow large time steps to be used
to solve the Newtonian equations of motion in MD simulations \citep{wang05}, as
well as multiple time step MD integrator \citep{2006MolPh.104..415M}.  Finally, with the
converged solution for the dipoles, we update the moments for the dipoles
stored in $\T^\mathrm{atom}$ and compute the dipole-dipole forces as well as
the dipole-$q^E$ forces. 

\section{Tests} \label{sec:tests}

To assess the quality of our implementation, both in terms of precision and
efficiency, we performed a number of accuracy and performance tests.  We
compare the force evaluations with those generated by the standard $O(N^2)$
implementation, which we refer to as a ``direct'' summation.  Tests computed
using a direct sum and involving polarized atoms use point dipoles and not
the dipole approximation from \secref{FMM and point induced dipole moment}.

We measure the relative error in the forces $\vec a_i$ of each particle via the quantity
\begin{equation}
    f^\mathrm{err}_i = {{|\vec a^\mathrm{fmm}_i - \vec a^\mathrm{dir}_i|} \over {|\vec a^\mathrm{dir}_i|}}
    \label{eq:error}
\end{equation}
We chose to compare the force rather than potential as the force is the
directly integrated quantity and because the potential is smoother and thus
less likely to reveal any computational issues.  Where timings are reported,
the calculations were performed on a single core Intel Core~i7 1.7GHz CPU.  We
compiled \polaris{} with production-run settings using the Intel ifort~14.0.3
compiler with the -O3, -xHost, and -ipo options and SIMD vectorization enabled. 

Most of the tests were performed by considering different kinds of molecular
systems \textit{in vacuum}. For instance, we test the scaling properties of our
FMM implementation using non-hydrated subsets of the HIV-1 capsid system.
However, we also performed a series of tests with a protein complex solvated in
a coarse grained solvent box, according to the computational approach proposed
by the second author \citep{masella08,masella11}. In that approach, the solvent
is modeled by polarizable pseudo-particles, and both the solvent-solvent and
solute-solvent interactions are truncated for distances longer than  7~\AA{}
and 12~\AA, respectively. The algorithmic complexity concerning the
solute-solvent and solvent-solvent interactions therefore scales as $O(N_p)$ and
$O(N_s)$, where $N_p$ and $N_s$ are the number of protein atoms and solvent
pseudo-particles, respectively. Moreover, periodic boundary conditions are used
to maintain the solvent density within the simulation box. They are 
applied only to the solvent pseudo-particles, whereas the solute doesn't
interact with its own periodic images (i.e., the computation of electrostatic
solute-solute interactions corresponds to a free boundary condition
problem). Hence, during our tests, FMM is used only to compute intra-solute
interactions and, as only atoms are involved in long range forces, we will
ignore the solvent pseudo-particles in the discussions.  Note that, as already
mentioned in \secref{sec:Introduction}, to rectify the solvent model deficiencies
originating from long range solute-solvent force truncation, we recently
proposed a multi-level coarse-grained $O(N_p)$ approach allowing one to
minimize the solute-solvent truncation drawbacks, particularly in the case of a
solute presenting charged moieties \citep {masella13}. The solvated
protein tests discussed here  will demonstrate the numerical accuracy of a
solute(FMM)/solvent(coarse-grained) approach, whose complexity  is $O(N)$ and
which accounts for all the microscopic forces present in the system, regardless
of their range. 

\subsection{Sphere Tests} \label{sec:sphere}

We first tested the FMM force accuracy using a randomly generated sphere of
atoms. The sphere was generated by uniformly sampling 4096 points inside a
radius of 45~\AA{} with a minimum interparticle separation of 2~\AA.
An equal number of $q=\pm e$ charges were chosen such that the sphere would
be charge neutral. Each atom was also assigned a random $q^E = \pm e$
charge for the electric field, but again in equal number to ensure neutrality. 

\begin{figure}
    \plotone{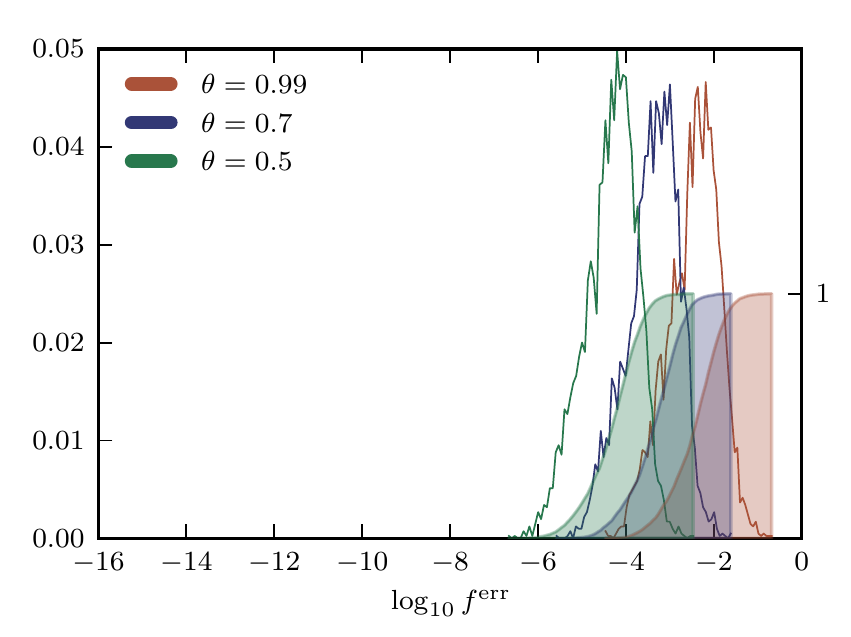}%
    \caption{The relative FMM error of the force on each atom of the random
        sphere compared with a direct evaluation. Polarization is disabled.
        Three values of the tolerance parameter $\theta=0.5, 0.7, 0.99$ are
        considered, from very accurate to less accurate.  The left vertical
        axis has been normalized such that the area under each curve is 1.
        Filled curves using the right vertical axis represent the cumulative
        histograms.}
    \label{fig:theta-test}
\end{figure}

In \figref{fig:theta-test} we show the relative error of the force with respect
to the direct evaluation. Polarization effects are disabled so that only the
Coulombic forces are considered. Three opening angles $\theta=0.5, 0.7, 0.99$
were chosen from more accurate to less accurate.  For $\theta=0.5$ about 80\%
of the atoms have errors less than $10^{-4}$ and all are less than $10^{-2}$,
as seen in the cumulative histograms. On our test machine, we measured the
average FMM computation time to be 0.23~s, 0.12~s, and 0.07~s for the three
$\theta$ values, respectively.  The direct summation required 0.06~s. For such
small systems the overhead introduced with FMM dominates the computation. 

\begin{figure}
    \plotone{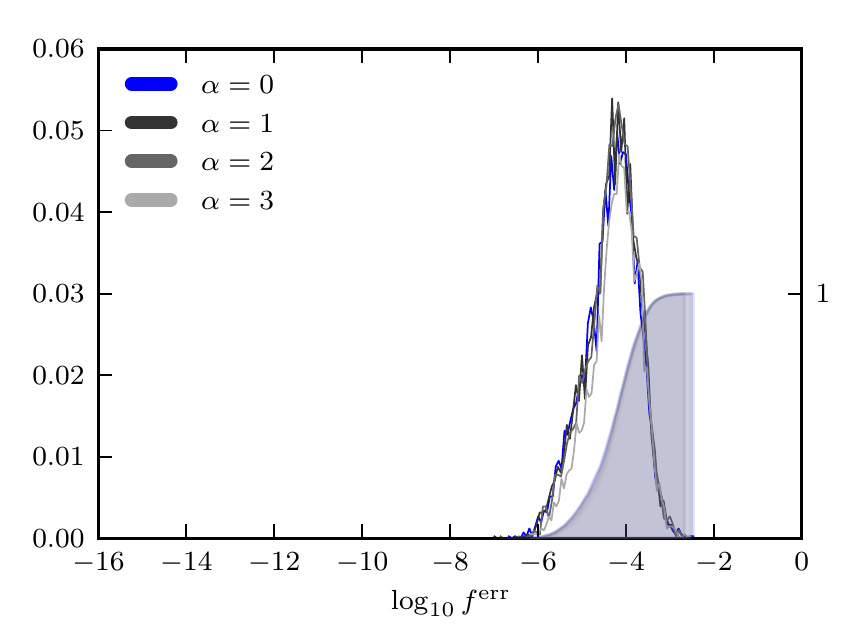}%
    \caption{The relative FMM error of the force on each atom of the random
    sphere compared with a direct evaluation. Polarization is enabled with
    where the polarization scalar $\alpha \in {0,1,2,3}$ has been assumed for
    all atoms.  The tolerance parameter is fixed at $\theta=0.5$. We highlight
    in blue the $(\alpha, \theta) = (0,0.5)$ case presented in
    \figref{fig:theta-test}. The left vertical axis has been normalized such
    that the area under each curve is 1. Filled curves using the right vertical
    axis represent the cumulative histograms.}
    \label{fig:pol-test}
\end{figure}

In \figref{fig:pol-test} we fix $\theta=0.5$, enable polarization, and recompute
the forces.  The polarization scalar for all atoms is taken to be $\alpha =
\{0,1,2,3\}$, for each test, respectively.  In the case of $\alpha=0$ the test
is equivalent to that shown previously in \figref{fig:theta-test}. 
In all cases there is no significant change in the error distribution.

\begin{figure}
    \plotone{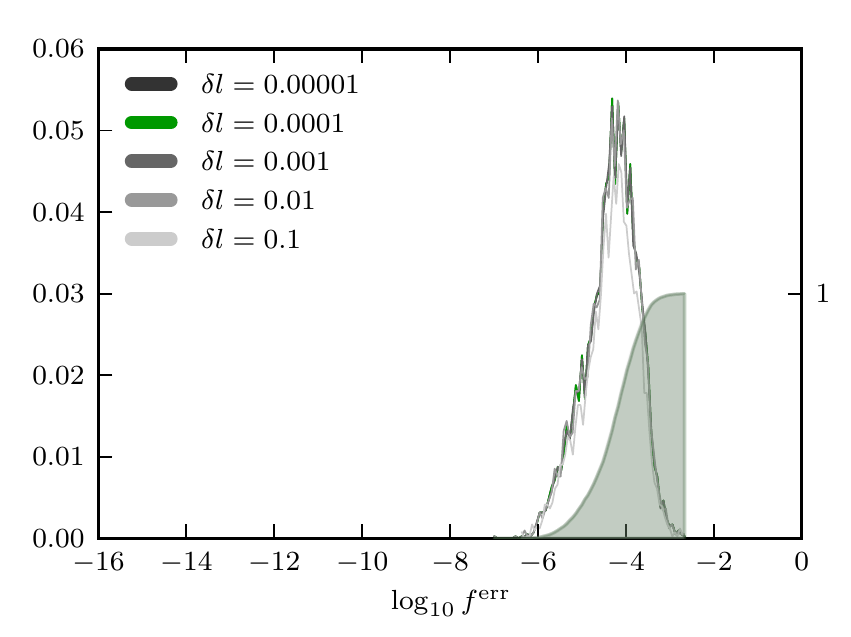}
    \caption{Force error distribution of FMM for the spherical test case
    without polarization considering different values of the dipole separation
    distance $\delta l = |\delta \vec l|$. Our choice for this approximation makes little
    difference over 5 orders of magnitude.  Here we assume an opening angle of
    $\theta=0.5$. We highlight in green the $(\alpha,\delta l)=(1,0.0001)$ case presented in
    \figref{fig:pol-test}.
        The left vertical axis has been normalized such that the area under
        each curve is 1. Filled curves using the right vertical axis represent
        the cumulative histograms.}
    \label{fig:pol-radius}
\end{figure}

We also considered the effect of our choice of $|\delta \vec l|$, the
separation distance for the point dipole approximation. Even though we tested
values over 5 orders of magnitude, from $10^{-5}$ to $10^{-1}$~\AA, as shown in
\figref{fig:pol-radius}, we again found no difference in the errors. We 
recommend a value of $|\delta \vec l| = 10^{-4}$ which is compatible with the
notion of a point dipole moment and is simultaneously large enough to avoid
any numerical issues.

A direct comparison of these parameters with other implementations is difficult
as many small details can affect the results. The most robust measure is to
demand a maximum error tolerance. We therefore chose for all of our future
tests the parameters $\theta=0.5, p=5, |\delta \vec l|=0.0001$~\AA{} to
ensure a maximum error of $10^{-2}$.

\subsection{Molecular dynamics of a solvated protein complex}

\begin{figure}
    \plotone{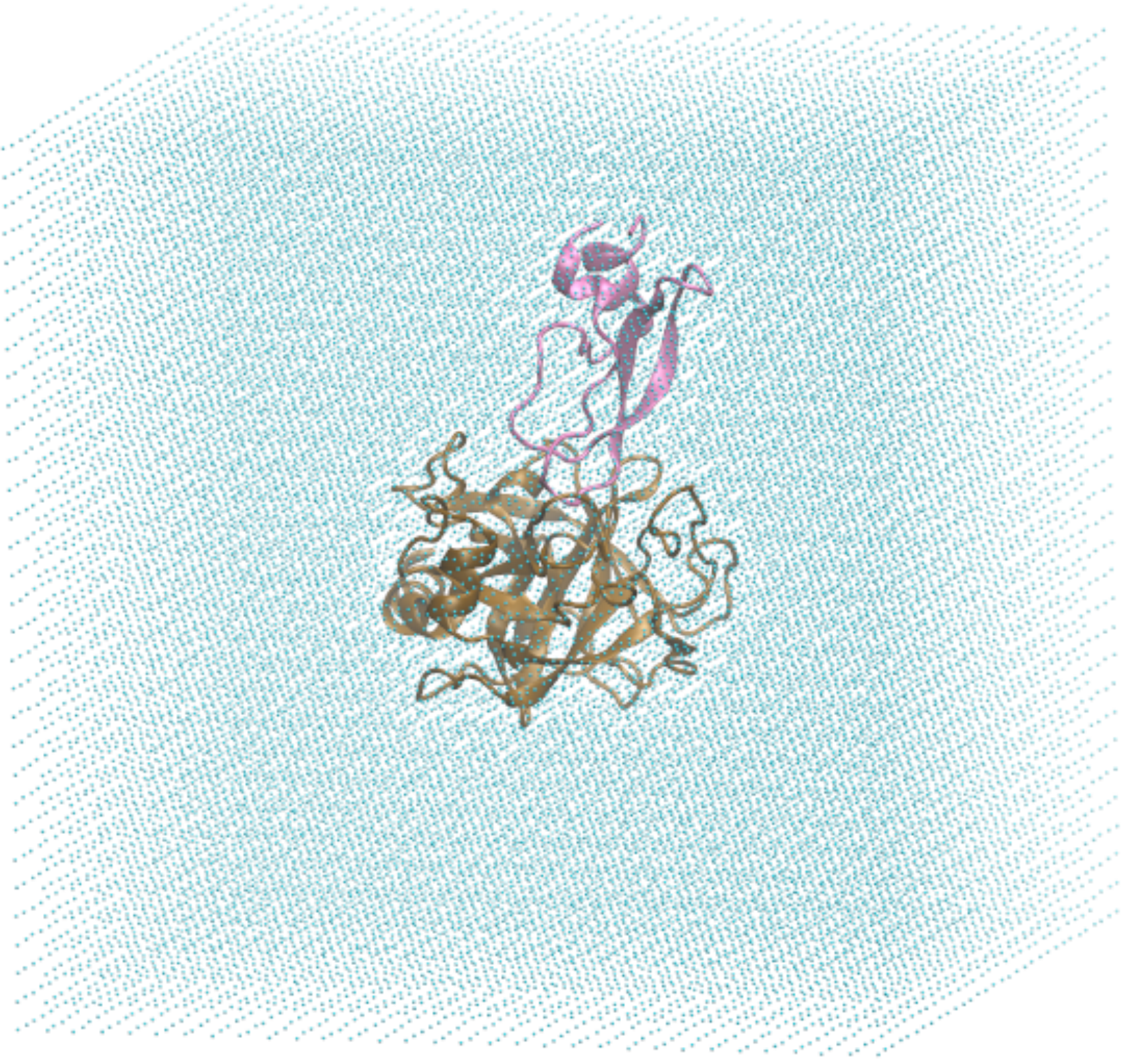}%
    \caption{The Pancreatic Trypsin Inhibitor-Trypsin complex (4,114~atoms)
    immersed in a pseudo-particle solvent (28,258~particles) non-periodic box.
    We perform six 1~ns simulations of the system to verify our FMM implementation.}
    \label{fig:2ptc}
\end{figure}

We used the Pancreatic Trypsin Inhibitor-Trypsin complex (PDB
label 2PTC \citep{Marquart:a22146,Berman01012000}) to test the
stability and accuracy of the FMM method used in conjunction with three MD
schemes over 1~ns. The protein complex is made of 2 non-bonded proteins and
consists of 4,114 atoms \citep{marquart83}. We immersed the system in a single
level solvent box made of 28,258 polarizable pseudo-particles (see
\figref{fig:2ptc}), according to the procedure described in \citet{masella11}.
As already mentioned, the computations concerning the electrostatic
interactions within the protein complex were performed in a non-periodic
environment  without any cut-off for the atom-atom interactions within the
solute and cut-offs at 12~\AA{} and 7~\AA{} for the solvent-solvent and
solute-solvent interactions, respectively.

Three different integrators were used with both the direct and FMM approaches.
The Newtonian equations of motion were solved using 1) the velocity Verlet
integrator with 1~fs time steps to handle non-bonded interactions, 2) the same
Verlet integrator with 2~fs time steps, and 3) the multiple time step (MTS)
integrator devoted to induced dipole moment based potentials
\citep{2006MolPh.104..415M}, where we used a time step of 1~fs for short range
non-bonded interactions and 5~fs for long range ones.  To handle the quickly
varying energy terms (those handling bonded atom interactions), we
systematically used a time step of 0.25~fs, regardless of the integrator.
Chemical bonds X-H and bending angles H-X-H were constrained to their initial
values using the RATTLE algorithm \citep{1983JCoPh..52...24A} with a
convergence criterion of $10^{-6}$~\AA. The induced dipole moments were solved
iteratively with a convergence criterion of $10^{-6}$ Debye per polarizable
center.  However, the iterations continue until the greatest difference between
two successive iterations of the induced dipole moment for a single polarizable
center is smaller than $20 \times 10^{-6}$ Debye. A temperature of 300~K was
monitored along the trajectories using the GGMT thermostat
\citep{2000JChPh.112.1685L} (with a coupling constant of 0.5~ps). The system
center of mass kinetic energy was subtracted before a simulation began.  Thus,
\textit{a priori}, we expect there to be no drift in the center of mass.
Lastly, based on the above spherical tests we use the FMM parameters
$\theta=0.5, p=5, |\delta \vec l|=0.0001$~\AA. All the systems were
equilibrated by performing 100~ps runs before starting the 1~ns production
runs. The values discussed below were extracted from these simulations.

We observed that the average number of iterations per time step needed to
converge the dipole moments using FMM or direct summation differed by about
2\%. The small difference originates mainly from the chaotic behavior of the
simulation trajectories, which explore different portions of the system
potential energy surface.

\begin{figure}
    \plotone{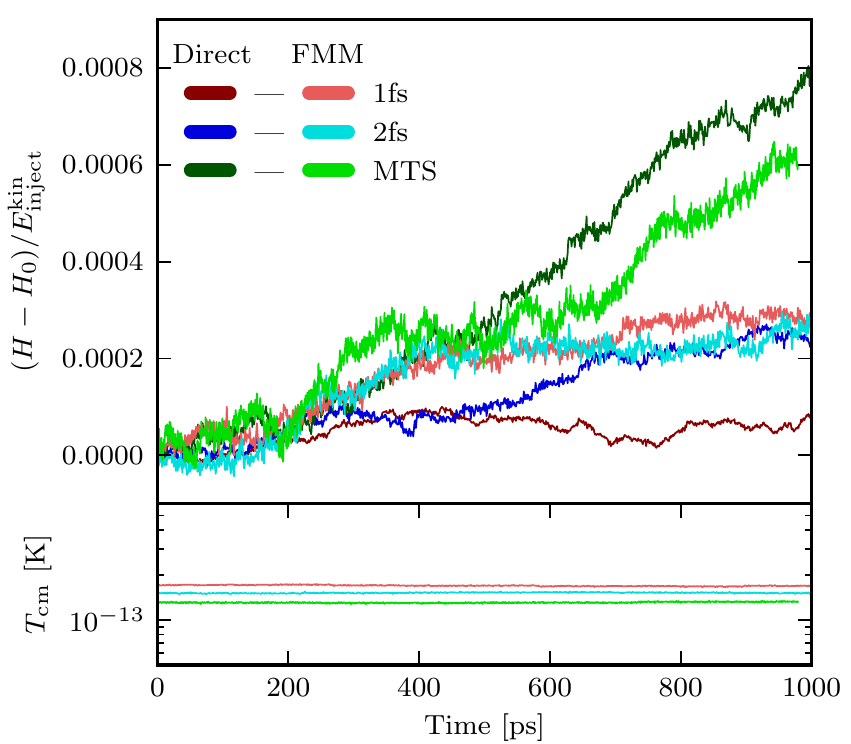}%
    \caption{\textbf{(Top)} The energy drift as a fraction of the total
    injected kinetic energy $E^\mathrm{kin}_\mathrm{inject}$. For the 2PTC
    protein simulation using fixed time step and multiple time step integrators
    $E^\mathrm{kin}_\mathrm{inject} =$ 63477 kcal/mol. The general behavior
    using FMM is similar to that of the direct sum computation.
    \textbf{(Bottom)} The kinetic energy  $T_\mathrm{cm}$ of the system center of mass
    with FMM, expressed in Kelvin. The small, nearly constant value for both fixed
    and multiple time steps demonstrates the expected momentum conserving nature of FMM.
    The 1~fs and 2~fs data has been offset vertically from the MTS run for clarity.}
    \label{fig:2ptc 1ns}
\end{figure}

The total Hamiltonian $H$ (which includes the GGMT thermostat components)
for the six 1~ns test cases are presented in the upper plot of
\figref{fig:2ptc 1ns}. The instantaneous total Hamiltonian has been adjusted by
the initial value $H_0$ and then scaled by the total injected kinetic energy
$E_\mathrm{inject}^\mathrm{kin}$ over the course of the simulation (here
$E_\mathrm{inject}^\mathrm{kin}=$ 63,477 kcal/mol). The initial 100~ps used to
equilibrate the system is not shown in the figure.

Because the dipoles are solved iteratively, a small energy drift is expected
\citep{JCP10.1063.1.1324708} and at the nanosecond scale we observed this mainly for the MTS
runs.  However, the drifts are small and all clearly under 0.1\%. Most
importantly, all the FMM runs behave consistently in terms of energy
conservation when compared with the direct summation based runs, exhibiting the
high precision level of the present FMM implementation. There may be small
inaccuracies due to the discontinuity introduced when a particle crosses a cell
boundary and transitions from a direct interaction to an FMM one, however,
these are largely dominated by errors in the iterative dipole scheme and do not
manifest themselves in the final results.

The lower plot of the same figure highlights the conservation of momentum of
the full system center of mass. As expected, there is essentially no change in
the value over the length of all simulations.

\subsection{FMM scaling and a comparison with PME}

We performed scaling tests using a series of five subset structures of the
mature HIV-1 capsid full system recently investigated in liquid water using a
PME approach and a non-polarizable standard pairwise force field
\citep{zhaonature2013}. The molecular structures considered here were extracted
from the PDB files 1VU\{4--9\}. Our base capsid subsystem is an hexamer of the
HIV-1 capsid protein. That subsystem contains 21,612 atoms (including
hydrogen). Larger capsid subsystems were generated by replicating this base
system up to a 5x larger system. Each base hexamer system interacts with the
others according to the interaction scheme of the HIV-1 capsid.  The largest
system considered is thus made of about $10^5$ atoms. All computations were
performed \textit{in vacuum} and represent a single energy point computation
using the same protocol to solve the dipole moments as for 2PTC and including
all short range forces in addition to the long range Coulomb and polarization
interactions.  

\begin{figure}
    \plotone{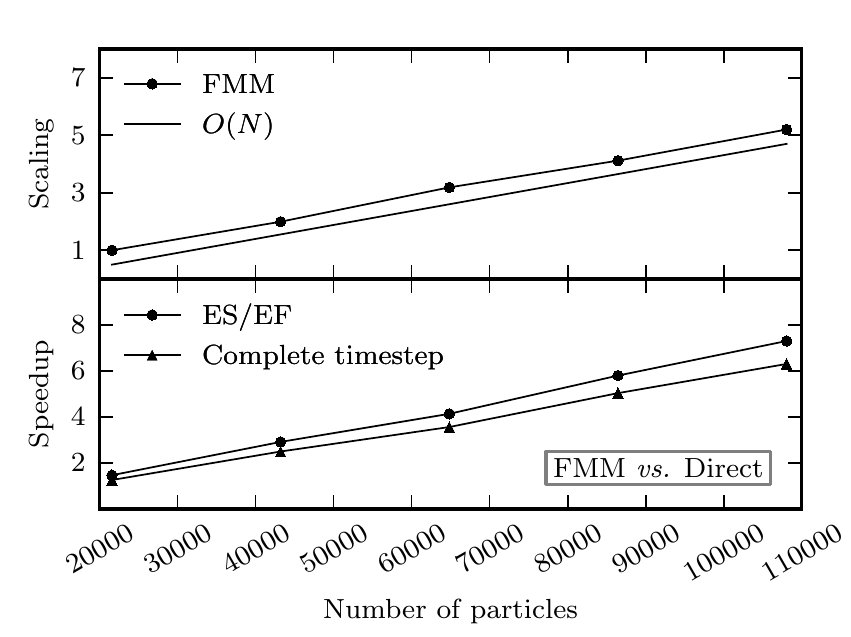}%
    \caption{%
    Performance properties of FMM with the HIV capsid subsystem for increasing
    system size.  \textbf{(Top)} The relative scaling of FMM (normalized by the
    smallest system), for the electrostatic/electric-field (ES/EF) calculation.
    The line labeled $O(N)$ represents perfect linear scaling.
    \textbf{(Bottom)} The speed up of FMM over direct summation when
    calculating either ES/EF or a complete timestep, which also includes the
    dipole convergence step. Detailed timing information can be found in
    \tabref{table:Timing}.}
    \label{fig:scaling}
\end{figure}

A plot of the FMM scaling and speedup compared with direct summation is shown
in \figref{fig:scaling}. For the scaling, we consider just the time to
perform the electrostatic force and electric field (ES/EF) calculation. This
isolates the FMM algorithm from other computations.  The FMM implementation
clearly scales linearly with the size of the system as indicated by the solid
line. The speedup plot compares FMM to a direct summation computation. Here, we
show the speedup both for the isolated ES/EF calculation and for a complete
timestep.  We find that FMM is about 7 times faster than the direct method for
the largest capsid system, although for the smallest system the two methods are
nearly comparable.  \tabref{table:Timing} provides the absolute timings used in
the plot. 

To efficiently simulate a molecular system (using free or periodic boundary
conditions), one may also consider a PME approach. To evaluate if an FMM
approach can be competitive to handle large molecular systems when accounting
for polarization, we also provide timing information for the capsid systems
when using the PME implementation in \polaris{}, which is based on the most
efficient PME approach proposed in \citet{1993JChPh..9810089D} and the
work of \citet{JCP10.1063.1.1324708} for induced dipole moment-based
force-fields.  The fast Fourier transform is performed using the FFTW 3.3.4
library \citep{FFTW05}, with SIMD vectorization enabled; most of the \polaris{}
loops involved in the PME energy/analytical gradient computations have also
been vectorized.  For consistency, we restricted the iterative scheme to solve
the induced dipole moments with 10 iterations, regardless of the computational
protocol used.  This is the upper bound of the number of iterations needed to
achieve the dipole convergence along a trajectory when considering the
convergence criteria used for the 2PTC system. 

We considered three typical grid sizes of 0.5~\AA{} (high resolution), 1~\AA{}
(medium resolution) and 2~\AA{} (low resolution). The PME direct energy terms
were truncated for inter-atomic distances larger than 12~\AA{}. When
considering the latter cutoff distance and a 1~\AA{} grid size, the PME
precision corresponds to truncating the direct and reciprocal standard Ewald
sums for terms smaller than 10$^{-8}$ \citep[][and references
therein]{2011PhR...500...43M}.  This is considered a high level of precision
for Ewald \citep[see e.g.,][]{JCP10.1063.1.1324708}.  The capsid systems were
set in their inertial frame of reference and the size of the enclosing boxes
correspond to the largest capsid coordinate in each dimension plus
$\sim$10~\AA{} (see \secref{sec:grid-dimensions}). We ensured that the
dimension values had small prime factors to take advantage of optimizations in
the FFT library.
 
For FMM, we used the parameters $\theta=0.5, p=5, |\delta \vec l|=0.0001$~\AA,
corresponding to a force accuracy where about 80\% of the relative error is
less than $10^{-4}$ and no more than $10^{-2}$. For a complete timestep, the
order of magnitude of the FMM timings is comparable to PME when using a grid
size of 1~\AA, but PME is faster than FMM by a factor of $\sim$3, regardless of
the capsid system size. PME is significantly faster, however, for the ES/EF
computations, up to a factor $\sim$10. The reason for the efficiency gain made
by FMM for a complete timestep is that while PME maintains the same grid for
the dipole convergence, FMM is able to build a more efficient tree using only
the dipole sites, which are about three times less than the total number of
atoms when using the polarizable force-field implemented in \polaris{}.  Given
further optimizations in our FMM implementation, we may reasonably expect the
FMM/PME gap to be smaller in the near future. The present FMM implementation
represents thus an interesting alternative to account for
electrostatic/polarization long range interactions when modeling large
molecular systems at the microscopic level. In particular, we may note here
that the FMM timings for an opening angle $\theta = 0.7$ match almost perfectly
the PME timings for a grid size of 1~\AA{} (results not shown in
\tabref{table:Timing}). Moreover, compared to using a high resolution PME grid
size of 0.5~\AA{}, the current FMM implementation is clearly faster by a factor
of a few.

\begin{table*}
{\scriptsize
\begin{tabular}{|r|ddd|ddd|ddd|ddd|ddd|}
\hline
\multicolumn{1}{|c|}{\mbox{}}             &
  \multicolumn{3}{c|}{Direct [s]}  &
  \multicolumn{3}{c|}{FMM [s]}     &
  \multicolumn{3}{c|}{PME 0.5\AA{} [s]}     &
  \multicolumn{3}{c|}{PME 1\AA{} [s]}     &
  \multicolumn{3}{c|}{PME 2\AA{} [s]}     \\
\multicolumn{1}{|c|}{No. Atoms}             & 
\multicolumn{1}{c}{ES/EF}                   & 
\multicolumn{1}{c}{Dipole}                  & 
\multicolumn{1}{c|}{Complete}               & 
\multicolumn{1}{c}{ES/EF}                   & 
\multicolumn{1}{c}{Dipole}                  & 
\multicolumn{1}{c|}{Complete}               & 
\multicolumn{1}{c}{ES/EF}                   & 
\multicolumn{1}{c}{Dipole}                  & 
\multicolumn{1}{c|}{Complete}               & 
\multicolumn{1}{c}{ES/EF}                   & 
\multicolumn{1}{c}{Dipole}                  & 
\multicolumn{1}{c|}{Complete}               & 
\multicolumn{1}{c}{ES/EF}                   & 
\multicolumn{1}{c}{Dipole}                  & 
\multicolumn{1}{c|}{Complete}               \\ 
\hline
21612  &  2.11 &    4.29 &    8.43 &    1.44 &    4.00 &    6.71 &    0.16 &   16.12 &   18.45 &    0.17 &    1.37 &    2.17 &    0.16 &    0.52 &    1.21 \\
43224  &  8.36 &   17.02 &   33.40 &    2.87 &    7.99 &   13.38 &    0.33 &   20.44 &   23.92 &    0.33 &    3.56 &    5.22 &    0.33 &    1.12 &    2.50 \\
64836  & 18.98 &   39.66 &   77.87 &    4.59 &   13.24 &   21.87 &    0.50 &   31.36 &   36.67 &    0.50 &    5.71 &    8.25 &    0.49 &    1.71 &    3.79 \\
86448  & 34.39 &   69.05 &  138.52 &    5.93 &   16.74 &   27.51 &    0.66 &   48.01 &   55.70 &    0.67 &    7.29 &   10.63 &    0.67 &    2.35 &    5.16 \\
108060 & 54.70 &  108.02 &  218.27 &    7.49 &   20.52 &   34.62 &    0.82 &   68.21 &   79.49 &    0.82 &   10.20 &   14.48 &    0.82 &    3.15 &    6.64 \\
\hline
\end{tabular}
}
\caption{Timing measurements (in seconds) for the capsid systems using direct summation,
FMM, and PME.  For each method we measured the time to complete the
electrostatic (ES/EF) calculation, the dipole convergence step, and the
complete timestep, which includes some additional computation. For the PME
approach, we used three typical grid sizes of 0.5~\AA{} (high resolution),
1~\AA{} (medium resolution) and 2~\AA{} (low resolution). The analysis was
performed on a single core of an Intel Core~i7 1.7GHz CPU.}
\label{table:Timing}
\end{table*}

\section{\label{sec:Conclusions}Conclusions}

We have implemented a momentum conserving version of the fast multipole method
in the molecular dynamics code \polaris{} and demonstrated the applicability of
FMM by combining it with state-of-the-art chemical models to compute, in $O(N)$
time, Coulomb forces and polarization interactions associated with
the induced dipole polarization method. Through a series of artificial and
real-world test cases---most notably, recent HIV-1 capsid subsystems from
\citet{zhaonature2013}---we have shown that we are able to achieve very good
accuracy by combining a multipole expansion to $5^\mathrm{th}$ order and an
opening angle tolerance criteria of $\theta=0.5$. During 1~ns MD simulations of
the 2PTC protein, we observe an energy drift of less than 0.1\% of the total
kinetic energy injected in the system and observe no change in the total
momentum---both consistent with standard $O(N^2)$ direct summation-based
simulations.

Performance measurements using the HIV capsid subsystems have demonstrated that
FMM can be $\sim$7~times faster than the direct summation method for systems of
about $10^5$ atoms. When accounting for polarization, the unoptimized FMM
routines are only a factor of a few slower than a PME approach using a 1~\AA{}
grid size and a factor of a few \emph{faster} than PME with a grid size of
0.5~\AA{}. Since FMM is expected to scale better than FFT-based approaches,
such as PME, particularly on forthcoming supercomputer architectures, it may
prove to be the method of choice for extremely large molecular dynamics
simulations of the future.

\begin{acknowledgments}

We thank Othman Bouizi (Intel Corp. and Exascale Computing Research Laboratory,
a joint \mbox{Intel/CEA/UVSQ/GENCI} laboratory) for his help in developing the
code \polaris. We also thank the two anonymous referees for their careful
reading of the manuscript and discussions that have greatly improved the text.
JPC would also like to thank W.~Dehnen and J.~Stadel for many useful and
insightful discussions. This work was granted access to the HPC resources of
[CCRT/CINES/IDRIS] under the allocation 2014-[6100] by GENCI (Grand Equipement
National de Calcul Intensif). We acknowledge access to the Stampede
supercomputing system at TACC/UT Austin, funded by NSF award OCI-1134872. 

\end{acknowledgments}

\appendix

\section{\label{sec:2PTC error distribution}Additional 2PTC tests} 

\begin{figure}
    \plotone{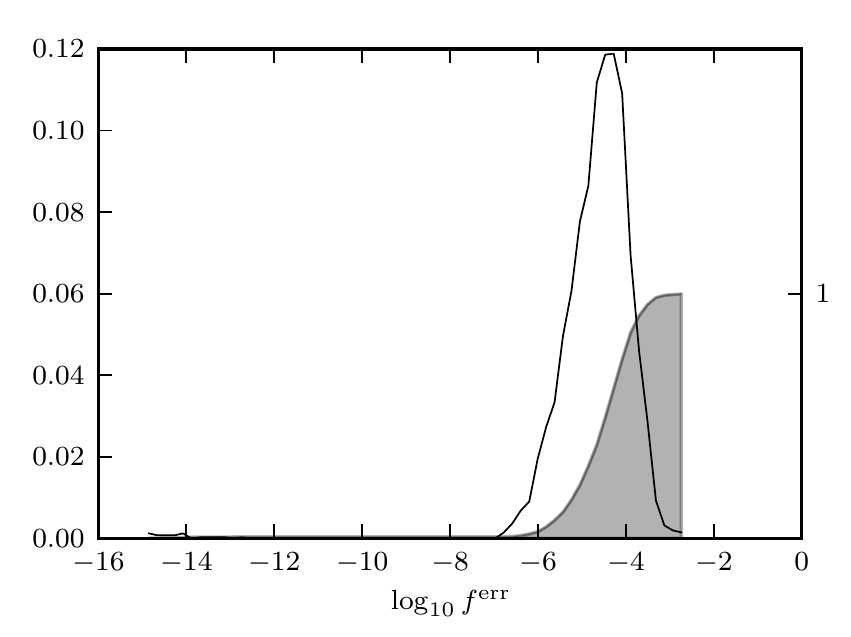}%
    \caption{Force error distribution of FMM after one time step of the 2PTC
    protein complex compared with a direct sum calculation.  The left vertical axis
    has been normalized such that the area under the curve is 1. The filled
    curve using the right vertical axis represents the cumulative histogram.}
    \label{fig:2ptc sgp}
\end{figure}

\begin{figure}
    \plotone{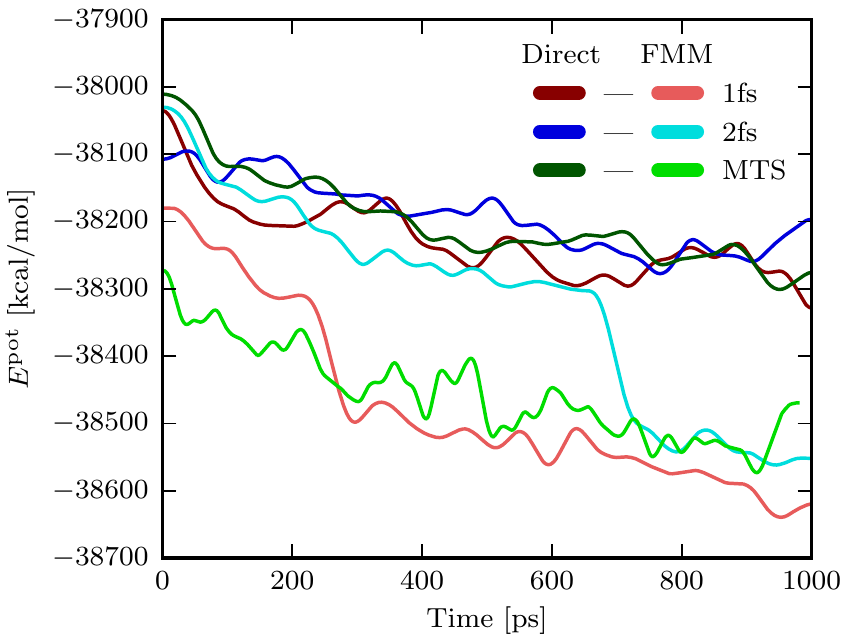}%
    \caption{The potential energy $E^\mathrm{pot}$ for the six 1~ns simulations
    of the 2PTC protein complex. The drift is due to the presence of 
    chaotic effects in the MD trajectories. The original data
    has been smoothed with a Hanning filter for clarity and to reduce the
    noise.}%
    \label{fig:2ptc Epot}
\end{figure}

In \figref{fig:2ptc sgp} we show the distribution of errors after a single
energy calculation of 2PTC. The distribution is remarkably similar to the
spherical tests with a peak in log-space just below $10^{-4}$.  In addition, we
present in \figref{fig:2ptc Epot} the temporal evolution of the system
potential energy $E^\mathrm{pot}$ over the course of the 1~ns simulations. For
each simulation, an initial 100~ps was used to relax the system using the same
procedure as the production phase (not shown in the figure). This results in
slightly different initial conditions for each run, as seen by a different
initial value of $E_\mathrm{pot}$. Therefore, this plot shows that different
portions of the system potential energy surface are explored along each
trajectory, and explains the difference in the number of iterations needed to
converge the dipole in direct sum and FMM simulations. The original data has
been smoothed with a Hanning filter for clarity and to reduce the noise.

\section{\label{sec:grid-dimensions}Grid sizes for PME}

The dimensions of the boxes, in which the capsid systems are embedded, are 
$112 \times 120 \times 128$,
$104 \times 136 \times 224$,
$112 \times 136 \times 320$,
$120 \times 140 \times 406$,
and 
$140 \times 140 \times 496$
for the PME simulations with a grid spacing of 1~\AA{}.  All dimensions in \AA.
For the grid spacing of 0.5~\AA{} the dimensions are twice these values, and
for the grid spacing of 2.0~\AA{} they are half of these values to maintain the
same physical box size.

\section{\label{sec:generating multipoles}Generating the multipole expansions}

Computing multipoles to order $p=5$ is a non-trivial task to perform by hand.
We have instead implemented a program in
\Python{}~\citep{Rossum:1995:PRM:869369} that symbolically generates the code
for all FMM operations. The equations are then symbolically simplified using
the \SymPy{} library~\citep{SymPy}. By eliminating common subexpressions we
see up to a ten fold decrease in operation count, although scaling still
behaves as $O(p^4)$.

\section{\label{sec:notation}Multi-index notation}

The multi-index notation simplifies the tensor representation in the derivation
of the FMM operations.  Following from property \eqref{mi1} addition and
subtraction operations are only valid for those tuples that have all $n_i \ge
0$.
\begin{eqnarray}
    \mi n           &=&      (n_x, n_y, n_z) \mathrm{\ where\ } n_i \ge 0       \label{mi1} \\
    |\mi n|         &\equiv& n_x + n_y + n_z                                    \\
    \mi n!          &\equiv& n_x!\; n_y!\; n_z!                                 \\
    \mi n \pm k     &\equiv& (n_x \pm k, n_y \pm k, n_z \pm k)                  \\
    \mi n \pm \mi m &\equiv& (n_x \pm m_x, n_y \pm m_y, n_z \pm m_z)            \\
    \vec r^{\mi n}  &\equiv& r_x^{n_x}\, r_y^{n_y}\, r_z^{n_z}
\end{eqnarray}

\bibliographystyle{aipauth4-1}  
\bibliography{v0}


\end{document}